\documentclass[fullpaper,twocolumn]{jpsj3} 

\usepackage{graphicx}
\usepackage{amssymb}
\usepackage{bm}
\usepackage{color}
\usepackage{ulem}

\newcommand{\bfit}[1]{\textbf{\textit{#1}}}

\newcommand{\unit}[1]{\rm #1}

\title{Ferrimagnetism Induced by Off-Site Coulomb Interaction \\ in an Itinerant Electron System}

\author{ 
Hirohito Aizawa
\thanks{aizawa@kanagawa-u.ac.jp, h.aizawa.phys@gmail.com} 
}

\inst{ Institute of Physics, Kanagawa University, Yokohama, Kanagawa 221-8686, Japan }

\abst{
Motivated by weak ferromagnetism (FM) in a $\tau$-type molecular conductor~($\tau$-MC), we examine its mechanism using a two-band extended Hubbard model. 
Applying the random phase approximation, we elucidate the uniform spin and charge fluctuations between unit cells in the presence of on-site and off-site interactions. 
Applying the mean-field approximation, we find the ordered state mixing with antiferromagnetism (AFM), weak FM, and charge ordering (CO) components in each unit cell: we classify this state as ferrimagnetism (FIM). 
We reveal the phase diagrams in the interaction and interaction--temperature spaces. 
The former shows that the off-site interaction induces FIM from pure AFM and the latter shows that lowering the temperature stabilizes FIM. 
To clarify the stabilization mechanism of the phases, we focus on the microscopic nature of the ordered states, including the band structure, Fermi surface, and density of states. 
We find that the FIM state is obtained from mixing features of AFM, CO, and FM; therefore, the emergence of FIM requires both the on-site and off-site interactions. 
Then, we discuss the effect of lowering the temperature and predict that the AFM gap assists the emergence of FIM based on AFM. 
This FIM state is possibly related to the observation of the weak FM in $\tau$-MC. 
}

\begin{document}
\maketitle

\section{Introduction} 
\label{Introduction}

The magnetic state in strongly correlated electron systems and its mechanism are attracting interest, and magnetism in the Hubbard model is actively being studied. 
In a half-filled Hubbard model with a bipartite lattice, antiferromagnetism~(AFM) or ferrimagnetism~(FIM) is stabilized by exchange interactions due to second-order perturbation~\cite{Lieb1989,Shen1994}. 
Studies of the mean-field~(MF) approximation and quantum Monte Carlo~(QMC) calculations have shown the stability of AFM~\cite{Hirsch1985,Hirsch1989,White1989,Imada1989,Furukawa1992,Furukawa1993} and the absence of ferromagnetism~(FM)~\cite{Rudin1985}. 
Numerous theoretical studies, including analytical and numerical calculations, on AFM applying the variational Monte Carlo~(VMC)~\cite{Inui1988,Giamarchi1991,Yokoyama2004,Yokoyama2013,Eichenberger2009,Yanagisawa2019}, many-variables VMC~\cite{Misawa2014}, cellular dynamical MF theory~\cite{Capone2006}, variational cluster approximation~\cite{Aichhorn2007}, Gaussian-basis Monte Carlo~\cite{Aimi2007}, path-integral renormalization group~\cite{Watanabe2004}, and constrained-path Monte Carlo~\cite{Chang2008,Chang2010} are still ongoing to reveal the mechanism of high-$T_{c}$ cuprates.

Several mechanisms have been proposed for the stability of FM. 
Nagaoka FM, which has been proposed to occur when one hole exists in a system with a large Coulomb repulsive interaction, is well known~\cite{Nagaoka1966,Tasaki1989}, and several numerical calculations support this FM~\cite{Kusakabe1992,Kusakabe1994na}. 
Flat-band FM, which is a characteristic one-body part of the Hamiltonian, is also well known~\cite{Mielke1991,Mielke1992,Tasaki1992,Mielke1993,Tasaki1994,Mielke1999,MielkeTasaki1993}, and numerical calculations on models with bipartite or second-nearest-neighbor~(NN) hopping have shown this form of FM~\cite{Kusakabe1994,Hlubina1997,Arita2000}. 
Moreover, FM stabilized by exchange interactions and a difference in the Hubbard-$U$ on the multiband model has been suggested~\cite{Hirsch1989exfm1,Hirsch1989exfm2,Tang-Hirsch1990exfm3,Hirsch1991exfm4,Strack1994,Wahle1998,Costa2016}. 
For a localized system, FIM caused by CO in a doped AFM Ising model has been suggested~\cite{Isoda2005}; this model shows an interesting relationship with the electronic state obtained in itinerant strongly correlated electron systems.

Recently, phenomena in which there is coexistence between spin ordering and charge ordering~(CO) have been devoted, and FIM induced by CO has been suggested. 
One of the suitable phenomena is the electronic state in manganite compounds~\cite{Tomioka1996,Chen1996,Bao1997,Kawano1997,Liu1998}. 
Several theoretical works indicate that CO shows interplay with magnetic orderings in one-dimensional~\cite{Garcia2000,Garcia2002,Neuber2006} and two-dimensional~\cite{Aliaga2001}  ferromagnetic Kondo lattice models, which are widely used to describe the manganite compounds. 
Density functional theory~(DFT) and Monte Carlo calculations give an analogous argument~\cite{Hamed2019}.

In low-dimensional molecular conductors~(MCs), CO and the charge-density wave~(CDW) are as important as magnetic properties~\cite{cr2004,jpsj2006,crystals2012,crystals2018}. 
In quasi-one-dimensional systems, various charge modulations have been found  experimentally~\cite{takahashi2006jpsj,dressel2012crystals} and  theoretically~\cite{seo2004cr,seo2006jpsj,yoshioka2012crystals,otsuka2014}, such as $4k_{\rm F}$ CO, $4k_{\rm F}$ CDW, $2k_{\rm F}$ CDW, and coexistence between CO and spin-Peierls states. 
For quasi-two-dimensional systems, richer CO patterns have been reported because of the magnitude and anisotropy of the electron-electron and electron-phonon interactions, such as checkerboard, horizontal, and stripe patterns~\cite{takahashi2006jpsj,tajima2006jpsj,shikama2012crystals,mori1995,seo2004cr,seo2006jpsj,mori2016,kuroki2006,kuroki2009,yoshimi2020}.

\begin{figure}[!htb]
\centering
\includegraphics[width=8.0cm]{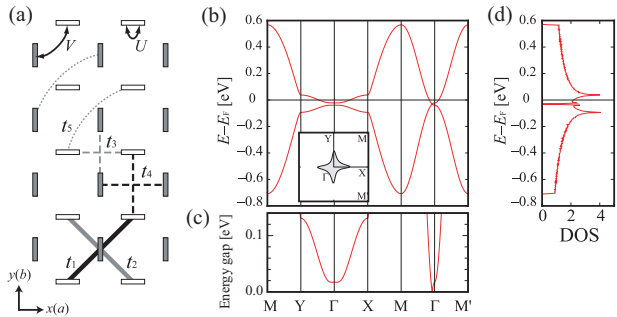}
\caption{(Color online) 
(a) Model of $\tau$-MC, where open and filled rectangles represent donor molecules. (b) Band structure, where inset is Fermi surface, (c) energy gap, and (d) DOS. 
}
\label{fig1}
\end{figure}

We have focused on the ordered state in a quasi-two-dimensional $\tau$-type MC~($\tau$-MC)~\cite{Zambounis1995,Papavassiliou1996}, where the $\tau$-type represents the molecular pattern with space group $I4_{1}22$ and the molecules are related to the glide symmetry operation. 
In our previous paper~\cite{Aizawa2014}, we reported a characteristic temperature dependence of the Seebeck coefficient and clarified its origin by applying the tight-binding model~[Fig.~\ref{fig1}(a)], which reproduces the first-principles band structure~[Fig.~\ref{fig1}(b)], where the hopping parameters are written later. 
The behavior of the Seebeck coefficient is related to the small energy gap near the center of the band structure. 
As shown in Fig.~\ref{fig1}(c), the valence and conduction bands are separated by an energy gap of 0.02 eV at the $\Gamma$ point, while there exists a band crossing along the $\Gamma$--M line. 
Figure~\ref{fig1}(d) shows that the density of states~(DOS) is large on the upper and lower parts from the center of the band structure, and a small energy gap separates the large DOS.

$\tau$-MC is formed with the chemical formula $\left[ D^{\left( 1+y \right)/2} \right]_{2} \left( A^{-1} \right)_{1+y}$~\cite{Papavassiliou1996,Aizawa2014,Terzis1991,Papavassiliou1992,Zambounis1995a,Papavassiliou1999,Papavassiliou2001,Papavassiliou2003,Yoshino2005}, where $D$ represents the donor molecule, $A$ is a linear anion such as AuBr$_2$, I$_3$, and so on, and $y$ is the anion content. 
Therefore, the average number of electrons on the donor molecule $n$ is written as $n=(3-y)/2$. 
Many MCs have $y=0$; however, the precise $y$ value is unclear for $\tau$-MC. 
Previous studies have found that the anion content $y$ can be estimated to be 0.75 from an elemental analysis~\cite{Papavassiliou1996} and 0.875 from an X-ray study~\cite{Konoike2003}. 
If $y=0.875$, which gives $n=1.0625$, $\tau$-MC is a slightly electron-doped system on a square lattice with half-filling. 
The system of $\tau$-MC can be considered as a half-filled square lattice with the carrier slightly doped or as a flat band with a dilute carrier doped to the band edge by focusing on the upper band.

Experiments on the magnetic properties in $\tau$-MC have shown a hysteretic magnetic state~\cite{Murata1998} and a weak FM~\cite{Konoike2001,Konoike2007} in $\tau$-(EDO-$S$,$S$-DMEDT-TTF)$_2$(AuBr$_2$)$_{1+y}$, where EDO-$S$,$S$-DMEDT-TTF is a donor molecule, ethylenedioxy-$S$,$S$-dimethylethylenedithio-tetrathiafulvalene. 
On the other hand, AFM has been observed in $\tau$-(P-$S$,$S$-DMEDT-TTF)$_2$(AuBr$_2$)$_{1+y}$, where P-$S$,$S$-DMEDT-TTF is an abbreviation of pyrasino-$S$,$S$-dimethylethylenedithio-tetrathiafulvalene~\cite{Konoike2007}. 
In addition, a theoretical study has suggested FM in a single-band Hubbard model on the upper band with a dilute carrier~\cite{Arita2000}. 
In our previous work~\cite{Aizawa2012}, we performed random phase approximation (RPA) analysis in a two-band extended Hubbard model consisting of a tight-binding model obtained from the DFT calculation of $\tau$-MC and an interaction term with an on-site and an off-site. 
The result showed that the spin and charge susceptibilities are maximum at the wavenumber vector $\bfit{Q}=\left( 0, 0\right)$. 
This behavior indicates the development of a uniform ordering whose periodicity is the same as that of the unit cell for the spin and charge channels, such as AFM, FM, and CO. 
To clarify the competition between the several ordered states, we must elucidate the arrangements of the spin and charge in the unit cell.

Given this background, we elucidate the electronic state and its mechanism in a two-band system on a square lattice with nearly half-filling, and note that a small band gap gives a property of a flat band with a dilute carrier. 
As the candidate model, we adopt a two-band extended Hubbard model of $\tau$-MC. 
In this model, several magnetic states are expected: 
(I) AFM originating from the nature of the half-filled band, 
(II) FM due to the flat band with the dilute carrier, and 
(III) a novel magnetic state. 
To verify the uniformity between the unit cell and its robustness, we apply the RPA. 
Since we cannot distinguish AFM and FM from the periodicity of a unit cell within this RPA approach, we use the MF approximation to clarify the spin and charge configurations in the unit cell. 
We show two phase diagrams to clarify the stability of the ordered state. 
Then, we elucidate a microscopic mechanism of the stabilization related to the several ordered states.

This paper is organized as follows. In Sect.~\ref{Method}, we describe the parameter set of the model and introduce the calculation methods. 
In Sect.~\ref{Results}, we present the uniformity obtained from the RPA; then, using the MF, we show the electronic state in the unit cell. 
We clarify that the electronic state is mixed with AFM, FM, and CO, and call this state FIM in this study. 
To examine the phase competition, phase diagrams in the interaction and interaction--temperature space are shown. 
To clarify the mechanism of the phase stability, we discuss the microscopic natures of the ordered states, which are the band structure, Fermi surface, and DOS. 
Finally, Sect.~\ref{Conclusion} reports our conclusions.

\section{Method} 
\label{Method} 

We adopt a two-band extended Hubbard model, which consists of the tight-binding model obtained from $\tau$-MC~\cite{Aizawa2014} and on-site and NN off-site interactions: 
\begin{eqnarray}
 H &=& \sum_{\bfit{k}, \alpha, \beta, \sigma}
  \varepsilon_{\alpha \beta}\left( \bfit{k} \right) 
  c_{\bfit{k} \alpha \sigma}^{\dagger} 
  c_{\bfit{k} \beta \sigma}
  \nonumber \\
  & & +\sum_{i \alpha} U n_{i \alpha \uparrow} n_{i \alpha \downarrow}
  +\sum_{\left< i \alpha; j \beta \right>}
  V_{i \alpha; j \beta}  n_{i \alpha} n_{j \beta}, 
  \label{hamiltonian1}
\end{eqnarray}
where $\varepsilon_{\alpha \beta}\left( \bfit{k} \right)$ denotes the one-electron energy dispersion with the momentum \bfit{k} between the sites $\alpha$ and $\beta$ in the unit cell, 
$U$ is the on-site interaction, $V_{i \alpha; j \beta}$ is the NN off-site interaction, 
$n_{i \alpha \sigma}$ is the number operator for electrons with the spin $\sigma$ at the site $\alpha$ in the unit cell $i$, and $\left< i \alpha; j \beta \right>$ represents the site pairs. 
We use the hopping parameters obtained from the first-principles calculation for $\tau$-MC~\cite{Aizawa2014}. 
These parameters are $t_{1}=0.161$, $t_{2}=-0.157$, $t_{3}=-0.011$, $t_{4}=0.021$, and $t_{5}=-0.003$ in \unit{eV} units; refer to the notation for $t_{i}$ in Fig.~\ref{fig1}(a), where the tight-binding model was used in our previous work and gives a clear understanding of the origin of the thermoelectric effect in $\tau$-MC~\cite{Aizawa2014}. 
In this study, the on-site interaction $U$, NN off-site interaction $V$, and temperature $T$ are variables.

To verify the uniformity of the electronic state between unit cells, we apply the RPA and obtain the spin and charge susceptibilities. 
The spin and charge susceptibility matrices $\hat{X}_{\rm sp}$ and $\hat{X}_{\rm ch}$, respectively, are given as 
\begin{eqnarray} 
 \hat{X}_{\rm sp}\left( \bfit{k} \right) 
 &=& \left[ \hat{I}-\hat{X}_{0}\left( \bfit{k} \right) \hat{U} \right]^{-1}
     \hat{X}_{0}\left( \bfit{k} \right) , 
 \label{chis} 
\\
 \hat{X}_{\rm ch}\left( \bfit{k} \right) 
 &=& \left\{ \hat{I}+\hat{X}_{0}\left( \bfit{k} \right) 
     \left[ \hat{U}+2 \hat{V}\left( \bfit{k} \right) \right]
     \right\}^{-1}\hat{X}_{0}\left( \bfit{k} \right) , 
 \label{chic} 
\end{eqnarray}
where $\hat{X}_{0}$ is the bare susceptibility matrix, $\hat{I}$ is the unit matrix, $\hat{U}$ is the on-site interaction matrix, and $\hat{V}$ is the NN off-site interaction matrix. 
Note that, in the two-band system, $\hat{U}$, $\hat{V}$, $\hat{X}_{0}$, $\hat{X}_{\rm sp}$, and $\hat{X}_{\rm ch}$ are all 2$\times$2 matrices. 
In the following, we diagonalize the spin and charge susceptibility matrices and concentrate on the largest eigenvalues, denoted $\chi_{\rm sp}$ and $\chi_{\rm ch}$, respectively. 
If the spin (charge) susceptibility diverges at $\bfit{Q}=\left( 0, 0\right)$, the spin (charge) configuration between unit cells is uniform. 
Because we cannot distinguish FM and AFM from the periodicity of a unit cell within this RPA approach, AFM emerges when spins are aligned antiparallel in the unit cell, while FM emerges when spins are aligned parallel.

To reveal the electron configuration in the unit cell, we apply the MF approximation. 
The MF Hamiltonian, which is obtained from self-consistent equations, gives the band structure of the ordered states. 
The stable electronic state is determined by comparing the free energy between the order parameters. 
We define the spin of the $z$-axis and the charge parameters on the site $\alpha$ as $S_{\alpha}^{z}=\left< n_{\alpha \uparrow} \right>-\left< n_{\alpha \downarrow} \right>$ and $C_{\alpha}=\left< n_{\alpha \uparrow} \right>+\left< n_{\alpha \downarrow} \right>$, respectively, where $\left< n_{\alpha \sigma} \right>$ is the average number of electrons with the spin $\sigma$ on the site $\alpha$. 
The order parameters of AFM, FM, and CO are 
\begin{eqnarray}
 S_{\rm AFM}^{z}=\frac{ S_{1}^{z}-S_{2}^{z} }{2}, 
 S_{\rm FM}^{z} =\frac{ S_{1}^{z}+S_{2}^{z} }{2}, 
 C_{\rm CO}     =\frac{ C_{1}    -C_{2}     }{2}. 
 \label{order-parameter}
\end{eqnarray}
In addition, we introduce a novel order parameter, in which AFM, FM and CO components are mixed and refer to the mixed state as FIM because only the $z$-axis spin is considered in this study.

Note that the interaction~(temperature) parameter tends to be smaller~(higher) than the realistic value because the RPA and MF calculations ignore several effects of the electron correlation. 
We, however, expect these approaches to give qualitative results for the competition between the different ordered states.

\section{Results} 
\label{Results}

Let us look into the electronic state obtained from the RPA and MF calculations. 
In the following, the band filling $n=1.0625$ corresponding to the anion content $y=0.875$ is adopted~\cite{Konoike2003}, and the system size is taken as 200$\times$200~(500$\times$500) in the RPA~(MF) calculation.

\subsection{RPA results: uniformity between unit cells and its robustness}
\label{RPA results}

\begin{figure}[!htb]
\centering
\includegraphics[width=8.2cm]{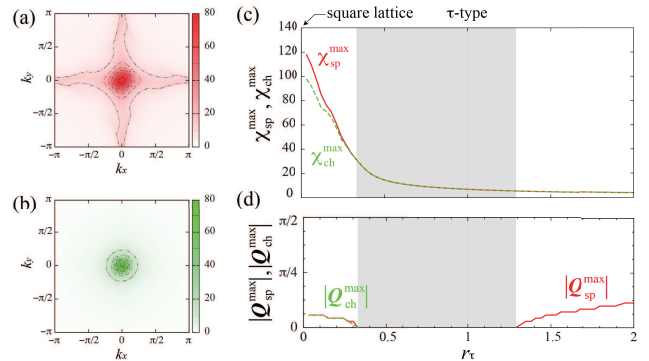}
\caption{(Color online) 
(a) Spin and (b) charge susceptibilities. 
(c) Maximum value and (d) its wavenumber of the spin and charge susceptibilities as a function of ratio $r_{\tau}$, where red solid (green dashed) curves represent spin (charge) part. 
}
\label{fig2}
\end{figure}

We verify the uniformity between unit cells and its robustness. 
Figures~\ref{fig2}(a) and \ref{fig2}(b), respectively show the contour plots of the spin and charge susceptibilities in our previous work,~\cite{Aizawa2012} where $U=0.35$ \unit{eV}, $V=0.0875$ \unit{eV}, and $T=0.002$ \unit{eV}. 
Both the spin and charge susceptibilities have similar maximum values at $\bfit{Q}=\left( 0, 0\right)$, which correspond to the uniform spin and charge fluctuations between unit cells. 
Our previous work~\cite{Aizawa2012}, however, has remained a possibility that the maximum susceptibility at $\bfit{Q} = \left( 0, 0\right)$ is accidental, because it is likely that the $\bfit{Q} \ne \left( 0, 0\right)$ state is stable in the square lattice with the incommensurate band filling.

To confirm the robustness of the uniformity, we introduce $r_{\tau}$, which is the ratio between $\tau$-MC and the pure square lattice. 
We use the hopping parameters such that $t_{i}'=t^{\rm sq}_{i} (1-r_{\tau})+t_{i} r_{\tau}$, where $t^{\rm sq}_{1}=0.16$ \unit{eV}, $t^{\rm sq}_{2}=-0.16$ \unit{eV}, and $t^{\rm sq}_{3}=t^{\rm sq}_{4}=t^{\rm sq}_{5}=0$ \unit{eV}. 
Figures \ref{fig2}(c) and \ref{fig2}(d) respectively represent the maximum value and its wavenumber $\left| \bfit{Q} \right|$ of the spin and charge susceptibilities as a function of the ratio $r_{\tau}$, where $U=0.22$ \unit{eV} and $V=0.055$ \unit{eV}. 
Introducing the property of $\tau$-MC to the square-lattice model suppresses the incommensurate behavior, although the incommensurate spin and charge susceptibilities are stable near the square lattice range. 
Then, the uniform arrangements of the spin and charge are widely present around the centered regime of $r_{\tau}=1$ corresponding to $\tau$-MC. 
Increasing $r_{\tau}$ develops the incommensurate spin susceptibility. 
Therefore, the RPA results indicate that the emergence of the uniform spin and charge fluctuations in the model of $\tau$-MC is robust between unit cells.

\subsection{MF results: ordered state in unit cell}
\label{MF results}

\begin{figure}[!htb]
\centering
\includegraphics[width=8.2cm]{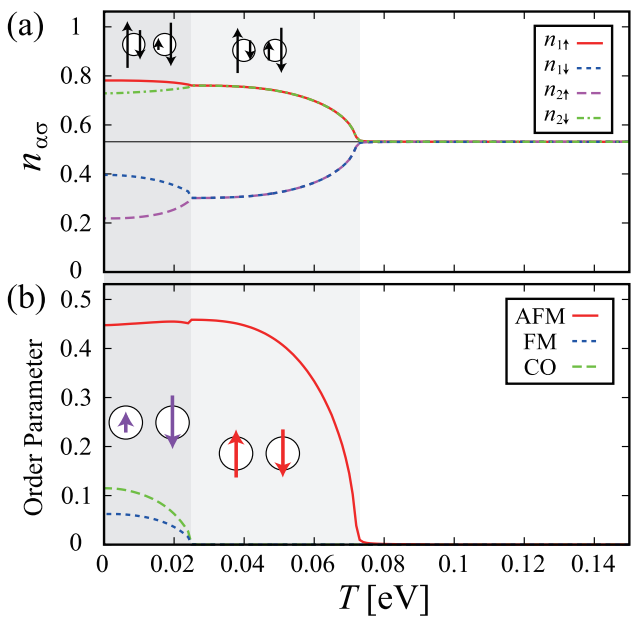}
\caption{(Color online) 
(a) Number of spin-$\sigma$ electrons at site $\alpha$ in unit cell and (b) components of order parameter as a function of temperature, where colored curves are represented in legend. 
The very light and light gray areas indicate the pure AFM and FIM, respectively. 
The insets represent schematic figures (a) of the electronic state and (b) order parameter at the sites. 
}
\label{fig3}
\end{figure}

To verify the spin and charge configurations in a unit cell, we apply the MF analysis to the two-band extended Hubbard model on $\tau$-MC. 
When we set the parameters $U=0.5$ \unit{eV} and $V=0.12$ \unit{eV}, we obtain the electron number $n_{\alpha \sigma}$ and the component of the order parameters as a function of the temperature as shown in Figs.~\ref{fig3}(a) and \ref{fig3}(b), respectively. 
Below the N\'{e}el temperature, which is approximately 0.07 \unit{eV}, the electron number of the spin on the same site differs, $\left| n_{1~(2) \uparrow}-n_{1~(2) \downarrow} \right| \ne 0$, namely, a pure AFM emerges; schematic pictures of the electronic state and order parameter are respectively shown in the insets in the very light gray area in Figs.~\ref{fig3}(a) and \ref{fig3}(b), where the arrow direction~(length) means the spin direction~(charge amplitude).

Moreover, upon lowering the temperature below approximately 0.025 \unit{eV}, the electron number $n_{1 \uparrow}$~($n_{1 \downarrow}$) differs from $n_{2 \downarrow}$~($n_{2 \uparrow}$) as shown in Fig.~\ref{fig3}(a). 
Figure~\ref{fig3}(b) shows that the weak components of FM and CO mix with the AFM component in the ordered state. 
To avoid the energy loss from the competition between the interaction and the kinetic energy, 
the electrons whose density on each site slightly deviates from one are unevenly arranged. 
In the AFM phase, the charge amplitude is equivalent; thus, the two arrows have the same length. 
At lower temperatures, however, the arrow lengths differ, and the weak components of FM and CO are mixed with the AFM component. 
Regarding the mixed ordered state as FIM, we find a phase transition from the pure AFM to FIM in the model of $\tau$-MC.

\begin{figure}[!htb]
\centering
\includegraphics[width=8.2cm]{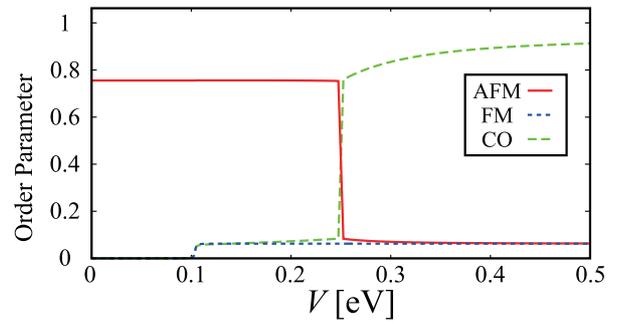}
\caption{(Color online) 
Components of order parameter as a function of NN off-site interaction $V$, where colored curves are described by legend of Fig.~\ref{fig3}(b). 
}
\label{fig4}
\end{figure}

Since the weak component of CO is mixed in FIM, the off-site interaction is expected to be important. 
Thus, we examine the effect of the off-site interaction. 
Figure~\ref{fig4} shows the components of the order parameter as a function of the NN off-site interaction $V$, where $U=1.0$ \unit{eV}. 
Upon increasing $V$, the weak components of FM and CO appear above approximately $V=0.1$ \unit{eV}; accordingly, we refer to FIM based on AFM as FIM$_{\rm A}$. 
If $V$ is larger than 0.25 eV = $U/4$, the dominant component changes from AFM to CO. 
We refer to the CO-based FIM as FIM$_{\rm C}$. 
On the basis of the results in Figs.~\ref{fig3} and \ref{fig4}, it is necessary to clarify the phase competition between AFM, FM, CO, and FIM.

\subsection{Phase diagram}
\label{phase diagram}

\begin{figure}[!htb]
\centering
\includegraphics[width=8.2cm]{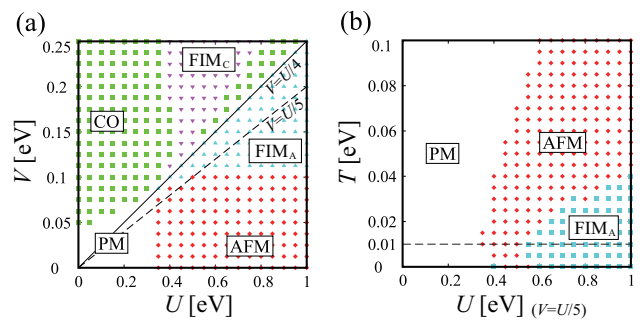}
\caption{(Color online) 
Phase diagrams in (a) interaction ($U$--$V$) space at $T=0.01$ \unit{eV} and (b) interaction--temperature ($U$--$T$) space at $V=U/5$, where black solid line in (a) indicates $V=U/4$ and black dashed lines correspond to parameters fixed in other phase diagram. 
}
\label{fig5}
\end{figure}

Figure~\ref{fig5}(a) shows a phase diagram in the $U$--$V$ space at $T=0.01$ \unit{eV}. 
Pure AFM~(CO) appears at large $U$~($V$) and small $V$~($U$), then FIM emerges around the regime satisfying the line $V=U/4$. 
Focusing on the regime of large $U$,  increasing the NN off-site interaction $V$ changes the pure AFM to the FIM$_{\rm A}$ phase; moreover,  when $V$ is larger than $U/4$, FIM$_{\rm A}$ gives way to FIM$_{\rm C}$. 
On the other hand, focusing on CO, the effects of $U$ and $V$ are exchanged, so increasing $U$ induces FIM$_{\rm C}$ from the pure CO phase. 
Figure~\ref{fig5}(a) gives a clue to clarifying the effect of the interaction on the stabilization mechanism of FIM, AFM, CO, and FM. 
In particular, it is necessary to distinguish the difference in mechanism between the FIM$_{\rm A}$ and FIM$_{\rm C}$, which we discuss later.

Figure~\ref{fig5}(b) shows the phase diagram in the $U$--$T$ space, where we fix the NN off-site interaction to $V=U/5$. 
When the on-site $U$ becomes larger than 0.4 eV, the AFM phase appears. 
Below the N\'{e}el temperature, FIM$_{\rm A}$ can be stable. 
Figure~\ref{fig5}(b) is associated with the observation of AFM~\cite{Konoike2007} and the weak FM in the low temperature regime~\cite{Murata1998,Konoike2001,Konoike2007} in the $\tau$-MC family.

\subsection{Mechanism of phase stability}
\label{mechanism}

\begin{figure}[!htb]
\centering
\includegraphics[width=8.0cm]{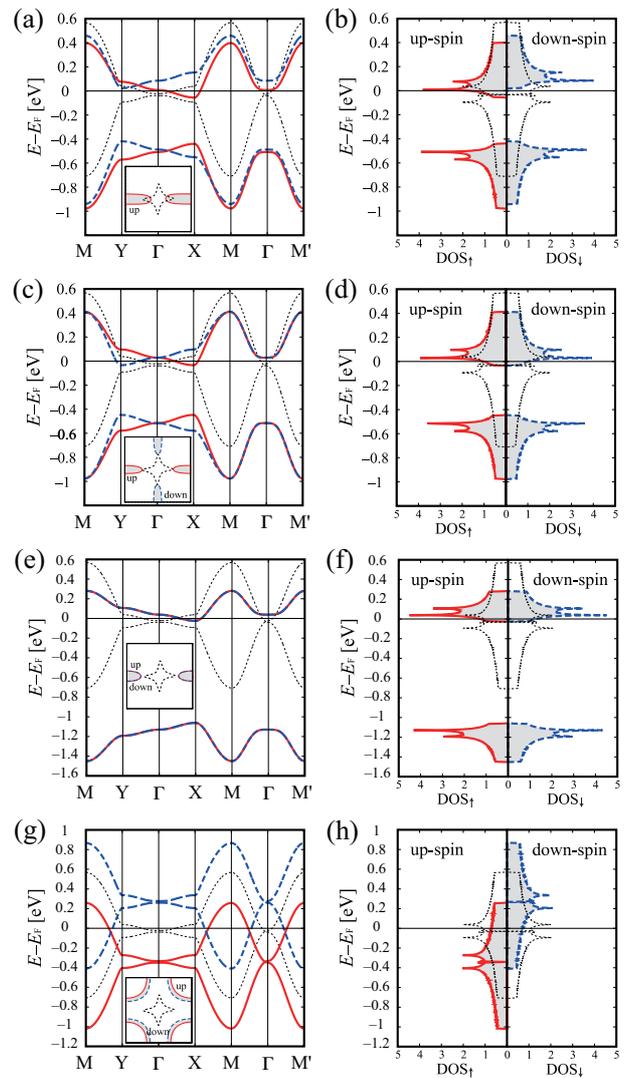}
\caption{(Color online) 
Band structure~(left panels), Fermi surface~(their insets) and DOS~(right panels) for FIM$_{\rm A}$~[(a) and (b)], AFM~[(c) and (d)], CO~[(e) and (f)], and FM~[(g) and (h)]. 
The red solid~(blue dashed) curves correspond to the up~(down) spin, and the black thin dotted curves represent the PM. 
}
\label{fig6}
\end{figure}

To reveal the stabilization mechanism originating from the interactions, first, we focus on the $U$--$V$ phase diagram~[Fig.~\ref{fig5}(a)]. 
In the following, we discuss the microscopic nature of FIM$_{\rm A}$, AFM, CO, and FM at $T=0.01$ \unit{eV}. 
Figure~\ref{fig6}(a), its inset, and Fig.~\ref{fig6}(b) represent the band structure, Fermi surface and DOS, in FIM$_{\rm A}$, respectively, where we set $U=0.8$ \unit{eV} and $V=0.15$ \unit{eV}. 
Features of the FIM$_{\rm A}$ state include the following three items: 
(I) the band structure splitting depending on the spins, 
(II) the anisotropy of the Fermi surface along the $\Gamma$-X and $\Gamma$-Y $k$ paths, and 
(III) the slight difference in electron number between the spins in the DOS. 
Keeping these features in mind, we discuss the stabilization mechanism of the FIM state based on the mixture of AFM, CO, and FM.

The band structure and Fermi surface for pure AFM are shown in Fig.~\ref{fig6}(c) and its inset, respectively, where $U=0.8$ \unit{eV} and $V=0.05$ \unit{eV}. 
An AFM gap of about 0.4 eV $\approx U/2$ is open, which can be understood as the degeneracy of the spin and the on-site $U$. 
The band dispersion splits in a manner depending on the spins along the $\Gamma$-X and $\Gamma$-Y $k$ paths; this behavior is of interest in relation to a recent theoretical work on the generation of a spin current in organic conductors $\kappa$-(BEDT-TTF)$_{2}X$~\cite{Naka2019}.  
In Fig.~\ref{fig6}(d), the DOS of the up spin is the same shape as that of the down spin, and the Fermi level exists at the same energy, namely, magnetization is absent. 
Since this system has incommensurate band filling, the Fermi level exists on the energy band. 
As shown in the inset of Fig.~\ref{fig6}(c), the Fermi surface of the up spin around the X point has the same size as that of the down spin around the Y point. 
The band splitting depending on the spins gives the characteristic nature of the spins in FIM$_{\rm A}$ shown in Fig.~\ref{fig6}(a).

Figure~\ref{fig6}(e) shows the pure CO band structure whose energy gap is approximately 1.2 eV$\approx 2(4V-U)$, where we set $U=V=0.2 \unit{eV}$. 
The band shows no magnetic property because of the degeneracy between the spins; naturally, the DOS of CO has the same shape between the spins as shown in Fig.~\ref{fig6}(f). 
The CO Fermi surface forms a pocket around the X point, while the DOS of CO is similar to that of AFM. 
Thus, CO is due to the magnetic degeneracy and anisotropy of the charge. 
This feature gives the anisotropy of the band structure and Fermi surface in the FIM$_{\rm A}$ state.

It is important to understand the weak magnetization in FIM$_{\rm A}$, although the pure FM is unstable in the whole phase diagram shown in Fig.~\ref{fig5}(a).  
The FM band shown in Fig.~\ref{fig6}(g) maintains the same shape as the PM band and shifts according to the spins, where $U=1.0$ \unit{eV} and $V=0.0$ \unit{eV}. 
As is well known, Fig.~\ref{fig6}(h) clearly shows that the imbalance of the electron depending on the spin gives FM. 
Similarly, the weak FM component in the FIM$_{\rm A}$~[Fig.~\ref{fig6}(b)] originates from the small difference in spin in the model of $\tau$-MC.

We summarize the stabilization mechanism of FIM$_{\rm A}$ by the following three items: 
(I) the band splitting depending on the spins comes from the feature of AFM, 
(II) the anisotropy of the band structure and Fermi surface originates from CO, 
and 
(III) the slight difference between the spins induces the weak FM. 
We expect that the pure FM and weak FM in FIM originate from the Stoner FM because the Fermi level exists around the large DOS. 
In the $U$--$V$ phase diagram, FIM emerges around the phase competition between AFM and CO~\cite{FIM_C}.

\begin{figure}[!htb]
\centering
\includegraphics[width=8.5cm]{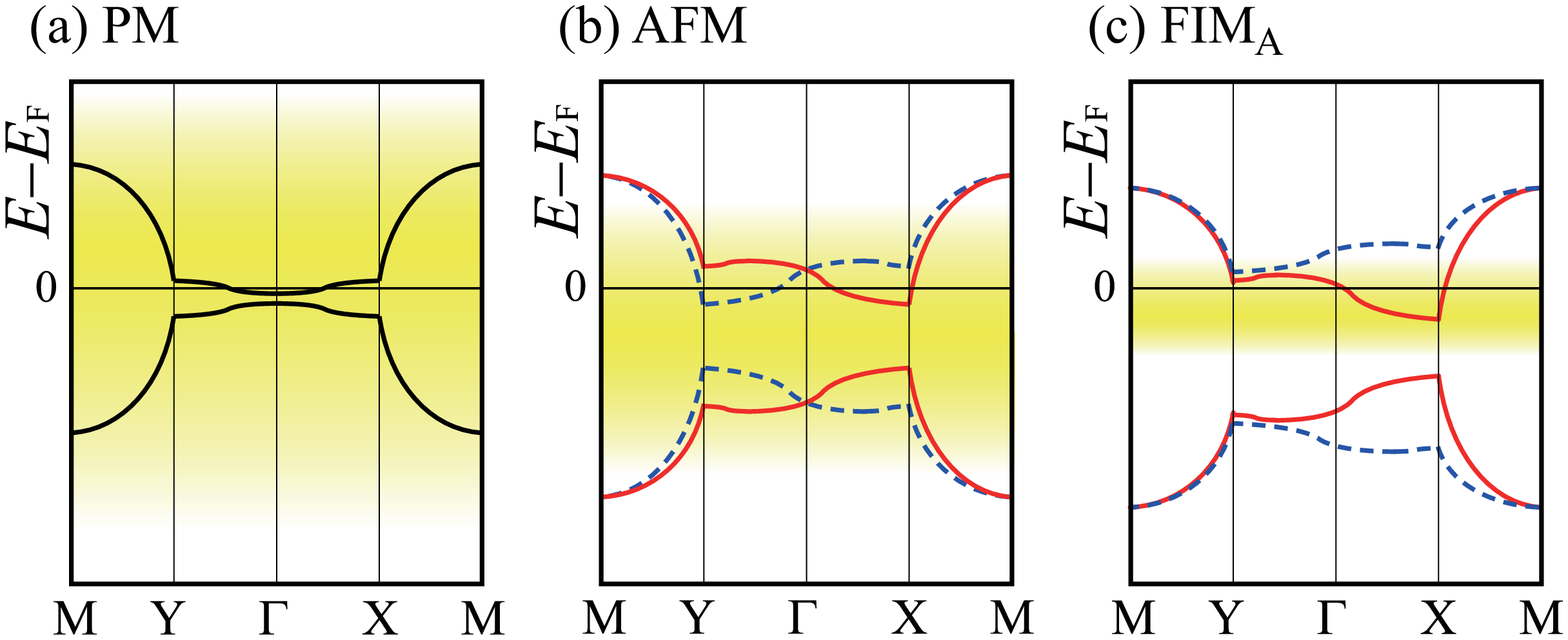}
\caption{(Color online) 
Schematic figures of relationship between band structure and temperature for (a) PM, (b) AFM, and (c) FIM$_{\rm A}$, where black solid curves represent noninteracting band dispersion and red solid~(blue dashed) curves represent band of up~(down) spin. 
The yellow regime schematically symbolizes the temperature effect from the Fermi level. 
}
\label{fig7}
\end{figure}

To understand the phase transition from AFM to FIM$_{\rm A}$ in the $U$--$T$ phase diagram~[Fig.~\ref{fig5}(b)], 
we discuss the schematic relationship between the band dispersion and the thermal effect as shown in Fig.~\ref{fig7}. 
In the high temperature regime, the PM band energy away from the Fermi level can affect the electronic state as shown in Fig.~\ref{fig7}(a), because of the thermal excitation. 
Upon lowering the temperature, the width of the temperature effect becomes small, as shown in Fig.~\ref{fig7}(b), and the energy range, which is approximately the band width, affects the phenomena. 
Therefore, the AFM state is dominant over the PM state because the model has a repulsive $U$ and approximately half-filling of the electron. 
At even lower temperatures, the thermal effect is suppressed, and the details near the Fermi level begin to affect the phenomena. 
Since the AFM gap has been open, the Fermi level exists at the bottom of the upper bands of both spins as shown in Fig.~\ref{fig7}(b). 
From the effect of the interaction on the phase stability, CO originating from the NN off-site interaction induces the anisotropic band dispersion such as that in Fig.~\ref{fig7}(c). 
Thus, when the off-site interaction is contained, the Fermi level exists on either spin band, and FIM$_{\rm A}$ can be stabilized at a lower temperature.

\section{Conclusions} 
\label{Conclusion}

We investigate the electronic state and its mechanism in a two-band system on a square lattice with nearly half-filling, focusing on the upper band, a flat band part with an effectively dilute carrier. 
As a candidate model, we adopt a two-band extended Hubbard model of $\tau$-MC in which a weak magnetization has been experimentally observed; the model has a small energy gap around the center of the band structure, which was a similar shape to the square-lattice band with a half-filled electron. 
We apply the RPA calculation and verify the uniformity between unit cells. 
Then, we examine the spin and charge configurations in each unit cell by applying the MF calculation.

From the RPA result, we find that the spin and charge susceptibilities have their maximum values at the wavenumber vector $\bfit{Q}=\left( 0, 0\right)$, which corresponds to the uniform configuration of the spin and charge arrangements between unit cells. 
We verify that even if the transfer integrals are shifted from the values of $\tau$-MC, the uniformity of the spin and charge configurations appears in a wide parameter regime.

We apply the MF calculation and find that lowering the temperature mixes AFM with CO and FM in the model exhibiting the on-site interaction $U$ and the NN off-site interaction $V$. 
Then, we indicate that increasing the off-site $V$ induces the AFM-based FIM state from the pure AFM state in a system with moderate $U$. 
Also,  further upon increasing $V$, the CO-based FIM state appears; note that both states have the weak FM component.

To understand the relationship between the interaction and the electronic state, we clarify the phase diagram in the interaction $U$--$V$ space. 
We find that FIM emerges around the regime satisfying the condition $V \approx U/4$ and that FIM$_{\rm A(C)}$ is dominant in the regime where $U/4$~($V$) is larger than $V$~($U/4$). 
Then, to clarify the robustness of FIM at low temperatures, we obtain the $U$--$T$ phase diagram and reveal that FIM$_{\rm A}$ can be stable below the N\'{e}el temperature.

To clarify the stabilization mechanism of FIM related to the other states, we present the band structure, Fermi surface, and DOS of FIM$_{\rm A}$, AFM, CO, and FM. 
We elucidate that the microscopic nature of FIM$_{\rm A}$ originates from the following features of AFM, CO, and FM: (I) the band splitting depending on the spins, (II) the anisotropy of the band, and (III) the emergence of the magnetization. 
Then, we schematically discuss the stability of FIM at low temperatures and predict that lowering the temperature suppresses the thermal effect. Subsequently, the detail of the band structure affects the phenomena.

This study on the magnetism in a strongly correlated electron system shows that the off-site interaction and lowering the temperature both induce the FIM state. 
The observation of the weak magnetization in the low temperature regime of $\tau$-MC has spurred interest in relation to the weak FM component, which is mixed in the FIM state in this study. 
To understand the mechanism of the magnetic state, analyses of other $\tau$-MCs with molecular replacement are necessary. 
Recently, a theoretical work has suggested that AFM energy band splitting generates a spin current in the AFM organic conductor $\kappa$-(BEDT-TTF)$_{2}X$~\cite{Naka2019}. 
The AFM band structure in this study has a similar shape, so its relationship with spin current physics is an interesting issue. 
In addition, the strongly correlated electron effect beyond the RPA and MF remains an interesting issue.

\section*{Acknowledgments}

The author acknowledges K. Kuroki for valuable discussions. 
This work has been supported by 
the Japan Society for the Promotion of Science KAKENHI Grants No. 16K17754, 
Grants-in-Aid from the Yokohama Academic Foundation, and 
the Science Research Promotion Fund from the Promotion and Mutual Aid Corporation for Private Schools of Japan.


\begin{thebibliography}{99}
\bibitem{FIM_C}
In FIM$_{\rm C}$, first, the pocket Fermi surface appears around the X point as shown in Fig.~\ref{fig6}(e). 
Then, upon mixing the AFM component, the band splitting depends on the spins as shown in Fig.~\ref{fig6}(c); therefore, FIM$_{\rm C}$ is stabilized.
\end{thebibliography}
\end{document}